# Irreversibility in Collapse-Free Quantum Dynamics and the Second Law of Thermodynamics

M. B. Weissman
Department of Physics
University of Illinois at Urbana-Champaign
1110 West Green Street
Urbana, IL 61801-3080

Abstract: Proposals to solve the problems of quantum measurement via non-linear CPT-violating modifications of quantum dynamics are argued to provide a possible fundamental explanation for the irreversibility of statistical mechanics as well. The argument is expressed in terms of collapse-free accounts. The reverse picture, in which statistical irreversibility generates quantum irreversibility, is argued to be less satisfactory because it leaves the Born probability rule unexplained.

**Introduction**

Irreversibility enters into fundamental physics through two main routes: the poorly understood quantum "measurement" process and the second law of thermodynamics. A relation between these two types of irreversibility has long been suspected. [1] The most prevalent view is that the underlying quantum formalism will remain time symmetric, with the manifest asymmetry of the measurement process ultimately resulting from the asymmetry of thermodynamics, in close analogy to the reasons why the retarded electromagnetic potential is more convenient than the advanced potential. [1] The point of this brief paper is to show that if proposals [2-6] that the measurement process results from non-linear decoherence processes which violate CPT symmetry [7] turn out to be correct, then the macroscopic behavior described by the second law would follow almost trivially as a consequence. The core of the argument will be that standard statistical irreversibility does not suffice to explain quantum probabilities, and thus leaves a requirement for some supplementary process. In contrast, a CPT-violating decoherence process would suffice for both purposes. An argument with ultimately the same logic, but based on a different physical picture, has also been made by Albert [8].

**Background**

It may help to review the irreversible aspect of quantum measurement, since it is not quite as widely appreciated as the irreversibility of thermodynamics. (See, e.g. [8] for a fuller description.) Although quantum field theory is itself fully CPT-symmetric (and most processes are actually T-symmetric), quantum phenomenology overall is highly asymmetric. The events we call "measurements" are ones in which a well-defined initial situation leads, via the purely deterministic evolution of the quantum state, to a superposition of two or more components representing macroscopically distinct outcomes. Observation always reveals only one of those outcomes, with the Born probability rule describing the likelihoods. Most importantly, this temporal evolution itself does not account for the Born probability rule, as has been repeatedly noticed, e.g. [6, 9-14]. The phenomena appear irreversible in that there are no known situations in which the current quantum state (as opposed to coarse-grained macroscopic properties) could have resulted from a number of macroscopically different prior states. Thus no time-reversed analog of a quantum measurement process has been observed. Therefore the phenomenon called quantum measurement is very strongly irreversible even though quantum field theory fully obeys CPT.

It is widely suspected that the irreversibility of measurement flows somehow from the irreversibility of thermodynamics.[1] Roughly speaking, this view holds that entanglement of the quantum state of the microscopic system being measured with the states of the large, statistically complex measurement apparatus results in permanent decoherence of the different macroscopic outcomes. For a state representing very large numbers of particles in two macroscopically distinct outcomes to somehow evolve to one in which the distributions of all coordinates of the entangled components overlap enough to interfere would require just the sort of coincidence which statistical mechanics in effect rules out.

There are two underlying problems with this view that the irreversibility of statistical mechanics leads to the irreversibility of quantum measurement. One is obvious and familiar, e.g. [1]- the origins of the irreversibility of statistical mechanics are themselves obscure. It is not clear why entropy systematically increases, i.e. why probabilities based on state-counting become increasingly applicable along only one direction in time. At the core of the problem lies the quantum Liouville thoerem: the entropy $Tr(-\rho \ln \rho)$, where $\rho$ is the density matrix, is conserved under Hamiltonian dynamics for isolated systems.

The second reason is less familiar but ultimately more serious. Decoherence of components of a quantum state does not, by itself lead to a unique macroscopic state but rather to "many worlds" [15, 16], i.e. decoherent components each including a version of each observer. If one does not wish to accept a many-worlds interpretation, an additional stochastic collapse process must be postulated (barring any sort of pre-collapsed Bohmian[17] interpretation), directly or indirectly, and such a process is not describable by a linear field theory. Attempts to describe such a process explicitly (rather than by mere verbalisms) make its intrinsic irreversibility explicit [2-5], and thus make the invocation of the irreversibility of statistical mechanics superfluous.

If, on the other hand, one accepts the possibility of a many-worlds interpretation, then one cannot independently insert the Born probabilities as an attribute of a non-existent collapse process. It has been argued repeatedly, e.g. [6, 9-14, 18] that simple decoherence does not give the correct Born probabilities or even *any* dependence of operationally defined probability on magnitudes of decoherent state components.

Many attempts to rescue the many-worlds picture have invoked hypothetical constructs beyond the quantum state, as reviewed by Saunders [19]. Others directly postulate Born probabilities or slightly weaker assumptions about the probabilities, as reviewed in [20], without examining whether these postulates are consistent with the operational tests of probability made via counting outcomes in the linear theory. To the best of my knowledge, only one attempt has been made to derive the Born probabilities from strictly linear many-worlds quantum mechanics without additional (and perhaps contradictory) constructs or axioms.[21] This attempt relies on the plausible argument that measure-dependent probabilities arise via anthropic constraints on the measure of any experiencable world in a background of partially decoherent components. However, in order to actually come close to Born probabilities one would need tight (and not yet justified) constraints on background decoherence rates for different physical outcomes.[21]

I have argued that some rather generic forms of non-linear collapse-free decoherence, a simpler non-random no-collapse version of processes hypothesized to give explicit state collapse[2, 22], result in Born probabilities[6], but only when the pointer operators driving the non-linear decoherence have an explicit time dependence, exponential growth, violating CPT symmetry. For the purposes of this paper, I do not wish to re-argue the question of whether any rational account can be made of actual quantum probabilities without modification of the Hamiltonian dynamics. It suffices to say that attempts to make a full, explicit, non-metaphysical description of quantum measurement, going from state-function to observation, *already* motivate and perhaps require postulating non-linear non-CPT processes to account for the probabilities, regardless of whether one insists on a unique outcome or puts up with many worlds. Such processes would obviate the need to invoke the irreversibility of statistical mechanics to explain the irreversibility of measurement. My point here will be that that such irreversible measurement processes would also *explain* statistical irreversibility, in that they *require* probabilistic predictions while making no such requirement for retrodictions, just as Albert has argued for collapse processes [8].

At first sight, the quantum and statistical irreversibilities contrast strongly. In a quantum measurement, a well-defined initial state (e.g. in Schrödinger's sadistic gedanken experiment) leads to any of several macroscopically distinct outcomes (say live cat or dead cat in a box). Statistical mechanics, in contrast, predicts that macroscopically distinct states (e.g. live cat or dead cat in a large box with air, water and food) will in the long run gradually evolve to members of a macroscopically unique equilibrium ensemble (e.g. an equilibrium gas of mostly $N_2$, $O_2$, $CO_2$ and $H_2O$ molecules).

This superficial contrast- increasing macroscopic diversity from quantum measurement vs. increasing macroscopic predictability from statistical equilibration- obscures an underlying similarity. In both the quantum measurement and the thermal equilibration cases, simple probability rules describe the future. I shall illustrate the underlying unity by considering the non-linear many-worlds view of quantum mechanics, because for this view the underlying rule for both types of irreversibility becomes identical: *the probabilities are determined by a simple count of possible decoherent quantum states*. In the collapse-based version of the same argument [8], different probability rules are invoked for the two types of irreversibility.

For any non-linear decoherence model, there is a corresponding non-linear collapse model consisting of the non-linear decoherence followed by purely random pruning of the decoherent branches. No experimental expectation value can change under such pruning, so the collapse picture would be observationally equivalent to the corresponding no-collapse picture. Therefore the argument will apply to either type of non-linear decoherence. However, we shall discuss briefly later the somewhat different global constraints on the collapse and no-collapse versions.

**A Toy Model**

The main idea can be illustrated with a toy model. Let us start off with a non-interacting gas of mesoscopic particles: small enough for quantum mechanical uncertainty relations to require that the wave functions spread out appreciably, but large enough to let the hypothetical non-linear decoherence processes be triggered occasionally for each particle. For further simplicity we can make the particles distinguishable.

Now let us put this gas out of thermal equilibrium in a one dimensional box of length L. For example, we can put all the particles at rest in the middle of the box, to within limits given by the uncertainty relations. The question then becomes whether linear quantum dynamics plus the irreversible non-linear decoherence processes would be sufficient to produce the equilibrium statistics for future particle positions, without making any spurious retrodictions for previous particle positions.

Let each particle have mass m, and let the state of the nth particle be represented by $\psi_n(x,t)$. We shall presume, imitating early attempts to produce explicit non-linear collapse theories[2], that $\psi_n(x,t)$ splits into decoherent components with typical spatial widths w at times of order $\tau$. Obviously, if this decoherence is not accompanied by collapse to one such component, one must postulate entanglement with additional physical variables ($\mu$) in order for decoherence to be possible, so we need to consider $\psi_n(x,\mathbf{v},t)$. [6]

Now let us consider the fates of one particle, prepared in an initial state with Gaussian profile and spatial width of order w. After time $\tau$, the resulting state has width of order $(w^2+(\tau\hbar/mw)^2)^{1/2}$. The non-linear decoherence then produces an ensemble of orthogonal $\psi_n(x, \mu ,t)$ states, which decohere due to having different values of the (unspecified) $\mu$. The variance $<(\delta x)^2>$ over the ensemble is equal to $<(\delta x)^2>$ evaluated immediately prior to the decoherence, i.e. $(w^2+(\tau\hbar/mw)^2)$. From each decoherent outcome, a new collection will then emerge over the next interval of duration $\tau$, again incrementing $\langle(\delta x)^2\rangle$ by an amount $(\tau\hbar/mw)^2$. The overall form of the sum of $|\psi_n^2(x,t)|$ over all elements of the ensemble will then obey a diffusion equation, with diffusion constant D of order $\tau(\hbar/mw)^2$). $\langle(\delta x)^2\rangle$ would then be a linearly increasing function of time if there were no walls. With walls, the solution to the diffusion equation of course simply approaches uniformity for times long compared to $L^2/D$ where L is a linear dimension of the box.

We now consider the operational meaning of that uniform-density solution. The uniform-density ensemble corresponds to a collection of decoherent outcomes in which, coarse-grained on a distance scale w or larger, each position for each particle is equally represented. Unless it could somehow be argued that the parameters which take on different values in different members of this ensemble were necessarily entangled with some observables outside the box, an observer who has not measured anything about the particle positions for a time longer than $L^2/D$ would *in principle* have no way of knowing which member of the ensemble she would ecpect to see upon making a measurement. Different 'she's' would see each possibility. Thus she should use the entire ensemble, counting each decoherent outcome once, to predict the probability of any measurement outcome. In other words, she should use standard statistical mechanics, restricted (in this particular simplified illustration) to a slightly peculiar microcanonical ensemble.

It is historically interesting that Graham[10] made a similar argument in reverse, but with somewhat hidden assumptions. After his famous remarks on the apparent irrelevance of measure to the operationally defined measured frequencies, he proposed that if results were read out through a macroscopic device on which expectation values were taken by unweighted averages over states, then measure must affect relative frequency. The current argument differs in that I claim that non-linear dynamics are required to make those unweighted averages give experimental expectation values.

**Some generalization**

The approach described above disposes of the conservation of Tr($-\rho \ln \rho$)) in the most direct (if purely hypothetical) possible fashion, by proposing that the dynamics are not purely Hamiltonian. The basic structure of our example has little to do with the particulars of the system or with the GRW guess about the particular form of the pointer operators. The non-Hamiltonian dynamics not only destroy the conservation of Tr($-\rho \ln \rho$)) but also lead to slight deviations from standard conservation laws. [5, 6] Pure states turn into ensembles, but the reverse process simply does not occur.

The relation between the preferred bases for 'measurement' and for statistical mechanics has often been considered before[23] [24] [25], However, with the exception of Albert's presentation of the effects of GRW-induced random fluctuations[8], discussions have generally assumed that the preferred basis arises by some sort of post-selection for state components supporting information-utilizing sub-systems, with purely linear Hamiltonian dynamics. It turns out that many of the considerations are not altered too much in the non-linear picture. There are some pointer operators specifying a preferred basis for measurement- whether due to non-linear contributions to the time dependence [2, 5] or to traces over an environment with which entanglement occurs via Hamiltonian dynamics. [25] Although the pointer operators and the Hamiltonian cannot commute (if anything is to actually happen), there are states which are *approximately* eigenstates of both. For example, in the GRW case these are the Gaussian wavepackets of width w. Such states provide a preferred basis in which it is possible to express the dynamics by a collection of probabilistic transitions occurring at rates small compared to energies/$\hbar$. The preferred basis for statistical mechanics thus consists of the sort of macroscopically definite states which we find as outcomes of quantum measurement processes. Thus within any such approach, in which ingredients of the time dependence operator include not only the Hamiltonian but also some non-linear functions involving pointer operators (in this GRW-like case, projections onto states of spatial width w [2]), the coarse graining involved in calculating a statistical density matrix is neither arbitrary nor purely anthropic.

To summarize the argument to this point, the most significant problem with an approach based on pure Hamiltonian dynamics is that it leaves little [21] or no [6, 9, 11, 13, 18] room to explain how the operationally-defined observed quantum probabilities arise. Our previously postulated solution of that problem [6] gives, at no *extra* cost, all the ingredients needed for a consistent statistical mechanics: preferred basis states and the requirement that statistics be used for determining what to expect in one time direction but not the other.

**Possible objections**

The proposal that thermodynamic irreversibility might arise from non-linear irreversible phenomena in modified quantum dynamics cannot be significant unless it is possible to make such modifications of quantum mechanics. Non-linear modifications of quantum dynamics have been considered long enough as possible resolutions of the measurement mystery to evoke serious objections. These include possible difficulties involving superluminal information transfer and, ironically, violations of the second law of thermodynamics.

Gisin [26] has pointed out that some non-linear modifications of quantum dynamics, e.g. ones of the type proposed by Weinberg [27], lead to violations of the causal limits on information transfer. Ferrero et al. [28] have argued that the key criterion is whether the time evolution of the probabilities (given axiomatically by the reduced density matrix $\rho$ in most approaches, and asymptotically by the same density matrix in the non-linear decoherence picture[6]) depends on space-like separated processes. Since the proposed non-linear decoherence[6] depends exclusively on *local* pointer operator projections, and induces only *local* decoherence, it would seem to meet the criterion for avoiding superluminal information transfer.

Albert has briefly discussed objections by Sklar and Pearle that under some circumstances the coarse-grained entropy can approach thermal equilibrium on shorter time scales than could plausibly be involved in any explicit quantum non-linear decoherence.[8] In the case of spin-echo experiments, for example, it is explicit that an apparent increase of spin disorder due to heterogeneity in precession rates can be mostly undone after a 180° rf pulse, and hence the actual entropy cannot have increased much. One cannot account for the apparent entropy increase by invoking hypothetical decoherence events which cannot yet have happened. However, there is nothing actually to be explained in such cases except for why we conventionally choose a coarse-grained basis within which an apparent entropy increase would be found if we neglect off-diagonal terms in the density matrix. Special initial states will be ones prepared to look simple in our conventional basis, because that is the basis corresponding to states of the apparatus with reasonably quasi-classical time dependence. That the (global) final outcome can never be a single such state would follow from the longer-term statistical irreversibility arising from quantum irreversibility.

Peres has argued [29] that non-linear modifications would actually allow violations of the second law- precisely the reverse of the effect discussed here. Consideration of why Peres' objection is not relevant helps clear up the connection between the two irreversibilities. Peres' argument requires the construction of an initial state which is the superposition of two orthogonal components which subsequently evolve under the non-linear dynamics into ones with positive scalar product. However, the irreversible dynamics proposed in the non-linear decoherence scheme [6] were constructed not to allow such a possibility, in order to avoid the possibility of (unobserved) recoherence events, i.e. time-reversed versions of the decoherence process. The new decoherent component acquires (in this toy model) a tag of new variables, specifying the time of the decoherence event. There are obviously no previously existing components with the same tag. It is satisfying that the description of a hypothetical dynamics intended to account for quantum probabilities and time asymmetry of measurement is driven to utilize just the sort of temporally expanding space of physically irrelevant variables needed to avoid problems with second-law violations.

**Collapse or plain decoherence?**

The recoherence issue raises a potentially important difference between non-linear collapse accounts of measurement[22] and statistics[8] and accounts involving only non-linear decoherence[6]. For non-linear collapse theories, the dynamics must be written in a manifestly time-irreversible fashion, with the absence of time-reversed measurement processes a basic postulate. For the non-linear decoherence approach, the absence of recoherence might be explained *either* by a manifest feature of the dynamical laws, which include CPT violations, *or* by the absence of nearby state components with which to recohere, as previously noted by Zeh [23]. In other words, a very low, or zero, density of such components (within whatever extension of ordinary Hilbert space might be required to represent the full states[6]) could account for the predominance of decoherence over recoherence, whose rate would be expected to be proportional to that density. Thus even if one accepts that the origin of statistical irreversibility lies in quantum irreversibility, some ambiguity would remain as to whether the explanation of both observed irreversibilities would ultimately be law-like or history-like[23], i.e. whether the space of states of unobserved variables has to be or happens to be sparsely occupied.

The non-linear decoherence and non-linear collapse (decoherence plus pruning) accounts also differ in a culturally interesting, if not experimentally testable, way. Traditionally, it has been assumed that statistical mechanical ensembles refer to possible outcomes, of which only one would be realized. There has, of course, been no unanimity about whether the possible outcomes of quantum measurement processes were all realized. Non-linear collapse pictures are unambiguous in treating these quantum measurement ensembles as hypothetical, with only one real outcome. In contrast, the non-linear decoherence picture unambiguously treats equilibrium statistical mechanical ensembles as being real long-time limits of the same real ensembles that appear in any many-worlds quantum interpretation. In this respect it shares with other many-worlds pictures any predictions for cosmology and biology which involve anthropic post-selection.